\documentclass[conference]{IEEEtran}
\usepackage[T1]{fontenc}
\usepackage[utf8]{inputenc}
\usepackage[english]{babel}
\usepackage{balance}
\usepackage{cite}
\usepackage{amsmath, mathtools}
\usepackage{amsfonts,amsthm,bm}
\usepackage{amssymb}
\usepackage{comment}
\usepackage{graphicx}
\usepackage{standalone}
\usepackage{dsfont}
\usepackage{mathtools}
\usepackage{color, colortbl}
\usepackage{soul}
\usepackage[normalem]{ulem}
\usepackage[acronym,shortcuts]{glossaries}
\usepackage{tikz}
\usepackage{arydshln}
\usepackage{bbm}
\usepackage{svg} 
\usepackage{algorithm,algorithmic}
\usepackage{balance}
\usepackage{adjustbox}

\usepackage{tikz}
\usepackage{pgfplots}
\pgfplotsset{compat=newest}
\pgfplotsset{plot coordinates/math parser=false}
\newlength\fheight
\newlength\fwidth
\usetikzlibrary{plotmarks,patterns,decorations.pathreplacing,backgrounds,calc,arrows,arrows.meta,spy,matrix}
\usepgfplotslibrary{patchplots,groupplots}
\usepackage{tikzscale}
\usetikzlibrary {patterns.meta}

\DeclareMathOperator*{\argmax}{\arg\!\max}

\newacronym{5g}{5G}{fifth-generation}
\newacronym{b5g}{B5G}{beyond-fifth-generation}
\newacronym{6g}{6G}{sixth-generation}
\newacronym{ai}{AI}{activation information}
\newacronym{amp}{AMP}{approximate message passing}
\newacronym{aoi}{AoI}{age-of-information}
\newacronym{awgn}{AWGN}{additive white Gaussian noise}
\newacronym{bs}{BS}{base station}
\newacronym{eccdf}{ECCDF}{empirical complementary cumulative distribution function}
\newacronym{cs}{CS}{compressed sensing}
\newacronym{cra}{CRA}{compressive random access}
\newacronym{csi}{CSI}{channel state information}
\newacronym{dt}{DT}{data transmission}
\newacronym{fifo}{FIFO}{first in-first out}
\newacronym{fsa}{FSA}{framed slotted-ALOHA}
\newacronym{gf}{GF}{grant-free}
\newacronym{gfeo}{GFEO}{greedy frame efficiency optimizer}
\newacronym{gmt}{GMT}{greedy maximum throughput}
\newacronym{harq}{HARQ}{hybrid automatic repeat request}
\newacronym{hmm}{HMM}{hidden Markov model}
\newacronym{iiot}{IIoT}{industrial internet-of-things}
\newacronym{iot}{IoT}{Internet-of-things}
\newacronym{isa}{ISA}{i.i.d. slot allocation}
\newacronym{kpi}{KPI}{key performance indicator}
\newacronym{lsfc}{LSFC}{large-scale fading coefficient}
\newacronym{lri}{LRI}{linear-reward-inaction}
\newacronym{m2m}{M2M}{machine-to-machine}
\newacronym{map}{MAP}{maximum a posteriori probability}
\newacronym{mdp}{MDP}{Markov decision process}
\newacronym{mimo}{MIMO}{multiple input-multiple output}
\newacronym{minlp}{MINLP}{mixed integer non-linear programming}
\newacronym{ml}{ML}{maximum likelihood}
\newacronym{mmpc}{MMPC}{min-max pairwise correlation}
\newacronym{mmtc}{mMTC}{massive \ac{mtc}}
\newacronym{mra}{MRA}{massive random access}
\newacronym{mtc}{MTC}{machine-type communications}
\newacronym{nr}{NR}{new radio}
\newacronym{noma}{NOMA}{non-orthogonal multiple access}
\newacronym{oma}{OMA}{orthogonal multiple access}
\newacronym{omp}{OMP}{orthogonal matching pursuit}
\newacronym{pam}{PAM}{pulse-amplitude modulation}
\newacronym{pb}{PB}{periodic beacon}
\newacronym{pdf}{PDF}{probability density function}
\newacronym{pia}{PIA}{partial information acquisition}
\newacronym{pima}{PIMA}{partial-information multiple access}
\newacronym{pomdp}{PO-MDP}{partially observable-Markov decision process}
\newacronym{ra}{RA}{random access}
\newacronym{rb}{RB}{reservation beacon}
\newacronym{rfid}{RFID}{radio-frequency identification}
\newacronym{rl}{RL}{reinforcement learning}
\newacronym{saloha}{SALOHA}{slotted ALOHA}
\newacronym{sb}{SB}{scheduling beacon}
\newacronym{sgfeo}{S-GFEO}{simplified \ac{gfeo}}
\newacronym{snr}{SNR}{signal-to-noise ratio}
\newacronym{tdma}{TDMA}{time-division multiple-access}
\newacronym{ura}{URA}{unsourced random access}
\newacronym{usd}{usd}{unit symbol duration}
\newacronym{urllc}{URLLC}{ultra-reliable low-latency communications}

\title{Minimum-Latency Scheduling For Partial-Information Multiple Access Schemes}

\author{Alberto Rech\IEEEauthorrefmark{1}\IEEEauthorrefmark{3}, Stefano Tomasin\IEEEauthorrefmark{1}\IEEEauthorrefmark{2}, Lorenzo Vangelista\IEEEauthorrefmark{1}, and Cristina Costa\IEEEauthorrefmark{4} \\ 
\small
\IEEEauthorrefmark{1}Department of Information Engineering, University of Padova, Italy.\\
\IEEEauthorrefmark{2}Department of Mathematics, University of Padova, Italy.\\
\IEEEauthorrefmark{3}Smart Networks and Services, Fondazione Bruno Kessler, Trento, Italy.\\
\IEEEauthorrefmark{4}S2N National Lab, CNIT, Genoa, Italy.\\
\footnotesize
\texttt{alberto.rech.2@phd.unipd.it, \{stefano.tomasin, lorenzo.vangelista\}@unipd.it, cristina.costa@cnit.it}
}
\begin{document}

\maketitle

\begin{abstract}
\Ac{pima} is an \ac{oma} uplink scheme where time is divided into frames, each composed of two parts. The first part is used to count the number of users with packets to transmit, while the second has a variable number of allocated slots, each assigned to multiple users to uplink data transmission. We investigate the case of correlated user activations, wherein the correlation is due to the retransmissions of the collided packets, modeling \ac{pima} as a \acl{pomdp}. 
The assignment of users to slots is optimized based on the knowledge of both the number of active users and past successful transmissions and collisions. The scheduling turns out to be a \acl{minlp} problem, with a complexity exponentially growing with the number of users. Thus, sub-optimal greedy solutions are proposed and evaluated. Our solutions show substantial performance improvements with respect to both traditional \ac{oma} schemes and conventional \ac{pima}. 
\end{abstract}

\begin{picture}(0,0)(0,-370)
\put(0,0){
\put(0,0){\qquad \qquad \quad This paper has been submitted to IEEE for publication. Copyright may change without notice.}}
\end{picture}


\glsresetall

\IEEEpeerreviewmaketitle

\section{Introduction}\label{sec:introduction}
The extremely demanding requirements imposed by \ac{b5g} \ac{iot} networks call for substantial improvements in multiple access techniques.   
While uncoordinated \ac{ra} paradigms such as \ac{noma} \cite{Saito13Non, Dai18A, Chen18Toward} and \ac{ura} \cite{Polyanskiy17A, decurninge2020tensor, rech2023unsourced} have gained significant attention in recent years, they require advanced pairing and power allocation techniques, as well as powerful channel coding and interference cancellation. Consequently, coordinated \ac{ra} techniques remain the preferred solutions for low-complexity multiple access. These solutions generally divide time into slots, each meant for the transmission of one packet.
\Ac{saloha}, wherein users transmit at the beginning of the first available slot after packet generation, is the simplest and most widely adopted coordinated \ac{ra} protocol. In case of collisions, it introduces a random delay before the re-transmission of the collided packets. Typically, the random delay follows the same statistics for all users, and the coordination is limited to slot synchronization \cite{Rivest87Network}. 
Collisions (leading to the accumulation of packets in user buffers) may induce correlated transmissions among users, which in turn increases collisions, but that can also be leveraged for indirect coordination in \ac{ra}. Recently, correlation-based schedulers have gained attention as potential breakthroughs for multiple access in \ac{iot}. Such schemes typically rely either on the knowledge of traffic generation statistics \cite{Kalor18Random}, or on its learning by \acp{hmm} \cite{Moretto21a} or reinforcement learning techniques \cite{Rech21Coordinated, Destounis19Learn}.
Additionally, \ac{cra} \cite{Wunder14Compressive, Choi20On}, wherein active users are first identified in the first sub-frame, upon the transmission of a preamble known at the \ac{bs}, has been shown to be effective in the adaptation of the instantaneous traffic condition.
However, such schemes lack scalability, as they need different preamble lengths in different traffic conditions, and typically require either \ac{mimo} receivers or multiple transmission steps to perform compressed sensing optimally.

In \cite{rech2023partial}, we introduced \ac{pima}, a semi-\ac{gf} coordinated \ac{ra} protocol, wherein time is organized into frames of {\em variable length}, each divided into two sub-frames. During the \ac{pia} subframe, active users (having packets in their buffers) send a signal to the \ac{bs}, which enumerates them through a compute-over-the-air approach  \cite{Goldenbaum13Robust}. Based on this knowledge, the \ac{bs} then assigns one slot to each user in the system for the transmissions in the \ac{dt} sub-frame, according to the optimization of the {\em frame efficiency}, i.e., the ratio between the expected number of successful transmissions and the frame length. 

In this paper, we propose a Markov model-based analysis of \ac{pima}, formalizing the case of correlated activations introduced in \cite{Rech2023Semi}. In particular, we model \ac{pima} as a \ac{pomdp} problem, which is solved by the Bellman equation. Then, to address the excessive complexity of the \ac{pomdp} solution, we propose a sub-optimal approach that determines the frame length iteratively, while performing the user scheduling, based on the frame efficiency metric.

The rest of the paper is organized as follows.  Section~\ref{sec:systemmodel}, introduces the packet generation process and the general \ac{pima} protocol framework. In Section~\ref{sec:POMDPmodeling}, we model \ac{pima} as a \ac{pomdp} and we derive the users' activation statistics. Section~\ref{sec:frameeff} presents two greedy scheduling solutions. Section~\ref{sec:numericalresults} show the numerical results of the proposed and existing solutions. Finally, in Section~\ref{sec:conclusions} we draw some conclusions.

\vspace{5pt}\noindent\emph{Notation.} Scalars are denoted by italic letters, vectors, and matrices by boldface lowercase and uppercase letters, respectively. Sets are denoted by calligraphic uppercase letters and $|\mathcal{A}|$ denotes the cardinality of the set $\mathcal{A}$. $\mathbb{P}(\cdot)$ denotes the probability operator and $\mathbb{E}[\cdot]$ denotes the statistical expectation.

\section{System Model}\label{sec:systemmodel}
According to the \ac{pima} setup of \cite{rech2023partial}, we consider the uplink of a multiple access scenario with $N$ users transmitting to a common \ac{bs}, with $N$ known at the \ac{bs}.

Time is split into {\em frames}, each comprising an integer number of {\em slots} for the \ac{dt} and an additional short time interval for the \ac{pia} phase. Each slot has a fixed duration $T_{\rm s}$. The whole system is assumed to be perfectly synchronized and each user transmits packets of the duration of one slot, at most once per frame.
In the following analysis, $\tau$ denotes a generic time instant, $t$ the frame index, and $\tau_0(t)$ the starting time of frame $t$. The same slot is in general assigned to multiple users for transmission.

When a slot is subject to collisions, all the packets in that slot are lost. This is the only source of communication errors.
We let $z_{n}(t) = 1$, if a successful transmission of user $n$ occurs at frame $t$, and $z_{n}(t) = 0$ otherwise. The case $z_n(t)=0$ includes also the event wherein user $n$ does not transmit at frame $t$. Successful transmissions are acknowledged by the \ac{bs} at the end of the current frame. Vector $\bm{z}(t) = [z_1(t), \ldots, z_N(t)]$ collects values for all the users. 

\subsection{Packet Generation and Buffering}
Packets generated at frame $t$ by user $n$ are stored in its buffer and transmitted at frame $t+1$, according to a \ac{fifo} policy. For user $n$, let $K_n(\tau)$ the number of packets in its buffer at time $\tau$. 
If $K_n(\tau) > 0$, the buffer of user $n$ is non-empty at time $\tau$, and the user is \textit{active}, instead, if $K_n(\tau)= 0$, its buffer is empty and the user is  \textit{inactive}. The total number of active users at $\tau_0(t)$ is denoted as $\nu(t)$, and the \textit{activation probability} of user $n$ at frame $t$ is 
\begin{equation}
\phi_n(t) = \mathbb{P}(K_n(\tau_0(t))>0).
\end{equation}
Note that, due to the presence of buffers and retransmissions, users' activations are in general correlated. Indeed, users colliding at frame $t$ will deterministically retransmit in the following frames, although in general in a different slot.
In the following, we assume that buffers are not limited. The traffic generation, also denoted as {\em packet arrival process}, at each user follows a Poisson distribution with parameter $\lambda$.

\begin{figure}
    \centering
    \includegraphics[width=\columnwidth]{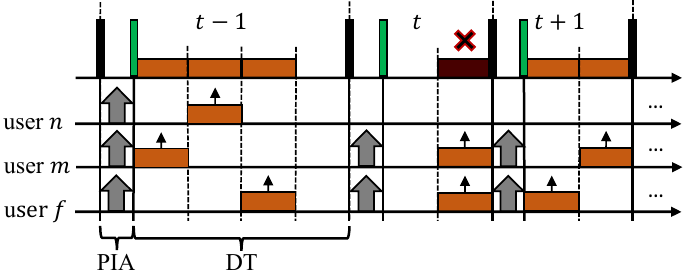}
    \caption{Example of \ac{pima} protocol operations and its frame structure. The \ac{rb} and the \ac{sb} are represented by black and green rectangles, respectively; grey arrows represent the uplink signals for the user enumeration at the \ac{bs} and data packets are represented by orange rectangles.}
    \label{fig:PIMAframe}
    \vspace{-5pt}
\end{figure}

\subsection{Partial-Information Multiple Access Protocol}\label{sec:protocol}
As shown in Fig.~\ref{fig:PIMAframe}, each frame is divided into two {\em sub-frames}, namely the {\em \ac{pia} sub-frame} and the {\em \ac{dt} sub-frame}. The \ac{pia} sub-frame is used to estimate (at the \ac{bs}) the number of currently active users. Based on this information and the past statistics of successful transmissions and collisions the \ac{bs} sets the duration (in slots) of the \ac{dt} sub-frame and assigns each user to one slot, for possible uplink data transmission. Such scheduling information is transmitted in broadcast at the end of the \ac{pia} sub-frame. 
We now briefly detail the \ac{pima} protocol, while in Section~\ref{sec:POMDPmodeling} we analyze its performance to derive policies for the slot assignment in the \ac{dt} sub-frame.

\vspace{5pt}\noindent\emph{Partial Information Acquisition Sub-Frame.}
At $\tau_0(t)$, the \ac{bs} transmits in broadcast the \ac{rb}, triggering the beginning of the \ac{pia} sub-frame and carrying the acknowledgments of correctly received packets in the previous frame. 
Moreover, \acp{rb} provide each user an estimate of its channel to the \ac{bs}, which is subsequently used for the user enumeration. Two approaches for the \ac{pia} sub-frame are described in \cite{Rech2023Semi} and \cite{rech2023partial}, based on a different knowledge of the channel state information at the user, and readers are referred to these papers for more details on the \ac{pia} sub-frame. Here we only recall that at the end of this phase, the \ac{bs} knows (possibly with errors) the number of active users, but not their identity. Then, the \ac{bs} schedules users' transmissions for the next \ac{dt} sub-frame.
Let $\bm{q}(t) = [q_1(t), \ldots, q_N(t)]$ be the \textit{slot selection vector}, collecting the slot indices assigned to each user.
The slot selection vector is transmitted by the \ac{bs} in the \ac{sb}, which triggers the beginning of the following \ac{dt} sub-frame. Note that the length $L_2(t)$ of the \ac{dt} sub-frame can be derived from $\bm{q}(t)$ as $L_2(t) = \max_{\substack{n}} q_n(t)$.
 
For the sake of simplicity, in the following, we assume fixed \ac{pia} sub-frame length $L_1$, comprehensive of both the enumeration task and the beacon transmissions, and assume an error-free user enumeration.

\vspace{5pt}\noindent\emph{Data Transmission Sub-Frame.}
In the \ac{dt} sub-frame, users transmit their buffered packets, according to the scheduling set by the \ac{bs} in the \acp{sb}.

\section{\ac{pima} Protocol Analysis}\label{sec:POMDPmodeling}

In this section, we describe the \ac{pima} protocol as a \ac{pomdp} and we introduce the latency as a penalty function to be minimized by the scheduler.

\Acs{pomdp} generalize the \acp{mdp} by combining such models with the key features of \acp{hmm}. 
An \ac{hmm} is defined by a Markov process, which is not directly observable (hidden process), and an observable random quantity depending only on the current state of the process (observation).  \Acp{pomdp} add the concepts of actions and penalties associated with the state transition-action couples, which are the fundamental features of \acp{mdp}. At each time period, the environment, represented here by the communication system, is in some state $i$. The agent, which is responsible to interact with the environment, takes an action, $a$, which causes the transition to state $j$ with probability $p_{i,j}(a)$. At the same time, the agent receives an observation $\bm{\beta}$  which depends on the new state of the environment $j$, and on the just taken action $a$, with emission probability $e_{\bm{\beta}} = \mathbb{P}(\bm{\beta}|a,j)$. Finally, when the system moves from state $i$ to state $j$ due to action $a$ at time $t$, the agent receives a penalty $P(t) = P(i, j, a)$, where $P(\cdot,\cdot,\cdot)$ is the penalty function. Then, the process is iterated.
The goal of the agent, at each time step, aims at minimizing its expected future penalty $\mathbb{E}\left[\sum _{t=0}^{\infty }P(t)\right]$. A discount factor may be also added to balance the impact of immediate and future penalties. 

Note that, differently from conventional \ac{mdp}, wherein the process state is known to the agent, in \ac{pomdp} the agent does not directly observe it, thus its actions are taken under uncertainty of the hidden state. 
However, by interacting with the environment and receiving observations, the agent may update its \textit{belief} in the hidden state, i.e., the state probability distribution based on the observations. Therefore, as the system evolves and the environment is observed, actions are taken with a more accurate estimation of the current state.

\subsection{\ac{pomdp} Model for The \ac{pima} Protocol}\label{sec4a}

In the following, we map the \ac{pima} protocol features to each of the discussed components of a \ac{pomdp}.
For the following analysis, we assume negligible error probability in the \ac{pia} sub-frame, i.e., \textit{perfect knowledge} of the number of active users, thus $\hat{\nu}(t) = \nu(t), \forall t$.

\vspace{5pt}\noindent\emph{Hidden States.} The hidden system state is the set of the arrival time of the packets in the buffers of all users. At time $\tau$, the state is $\bm{\mathcal{X}}(\tau) = \{x_{n, h}, \forall n, h\}$, where $x_{n,h}$ is the instant of the generation of the $h$-th packet, for all packets currently in its buffer. The state space of the \ac{pomdp} $\mathcal{Y} = \{\bm{\mathcal{X}}(\tau), \; \forall \tau\}$ has infinite cardinality, due to the infinite buffer length assumption. The buffer state evolves at each transmission attempt or packet generation, however, only the state probability at the beginning of the \ac{dt} sub-frame is relevant for scheduling.

\vspace{5pt}\noindent\emph{Actions.} The scheduling vector $\bm{q}(t)$ is the action performed by the agent \ac{bs} at frame $t$ based on its belief on the current system state. For simplicity, in the following, we denote the state at the end of the \ac{pia} sub-frame as $\bm{\mathcal{X}}(t)$. 

\vspace{5pt}\noindent\emph{Observations.}
Observations are acquired at the end of the \ac{pia} sub-frame of frame $t$ as
\begin{equation}\label{defos}
    \bm{\bm{\beta}}(t) = \{\nu(t), \bm{z}(t-1), \mathcal{C}(t-1)\} \in \mathcal{O},
\end{equation}
where set $\mathcal{C}(t)$ collects all slots wherein a collision is detected in frame $t$, and  $\mathcal{O}$ is the set of possible observations. Note that acknowledgment and collision vectors are referred to the previous frame, as we are at the beginning of the \ac{dt} sub-frame of frame $t$.  
The cardinality of $\mathcal{O}$ is finite, as only a limited number of combinations of successes, collisions, and number of active users may occur. 

\vspace{5pt}\noindent\emph{Conditional Transition Probabilities.}
The underlying Markov process evolves from state $\bm{\mathcal{I}}\in \mathcal{Y}$  at frame $t-1$  to state $\bm{\mathcal{J}}\in \mathcal{Y}$  with  posterior conditional \textit{transition probability}
\begin{equation}\label{transProb}
    p_{\bm{\mathcal{I}}\bm{\mathcal{J}}}(\bm{a}) = \mathbb{P}\left[\bm{\mathcal{X}}(t) =\bm{\mathcal{J}} |\bm{\mathcal{X}}(t-1) = \bm{\mathcal{I}}, \bm{q}(t-1)= \bm{a}\right]. 
\end{equation}
We observe that, by imposing the condition on the observations and the previous action, transitions occur only according to the packet generation and transmission processes.

\vspace{5pt}\noindent\emph{Emission Probabilities.}
For each state, different observations may be acquired in general. The emission probability of observing $\bm{\beta}'$ by visiting state $\bm{\mathcal{J}}$ is defined as
\begin{equation}
    e_{\bm{\beta}'}(\bm{\mathcal{J}}, \bm{a}) = \mathbb{P}[\bm{\beta}(t) = \bm{\beta}' | \bm{\mathcal{X}}(t) = \bm{\mathcal{J}}, \bm{q}(t) = \bm{a}].
\end{equation} 
\vspace{5pt}\noindent\emph{Beliefs.} We first observe that the information at the end of the \ac{pia} sub-frame of frame $t$ is both the observations and the actions up to that frame, i.e., 
\begin{equation}
\mathcal{E}(t) = \{\bm{\beta}(t),\bm{\beta}(t-1), \ldots, \bm{\beta}(0), \bm{q}(t-1), \ldots, \bm{q}(0)\}. 
\end{equation}
Now, the belief of state $\bm{\mathcal X}$ is the conditional probability of being in state $\bm{\mathcal X}(t)$ at frame $t$, given the information available at the \ac{bs}
\begin{equation}
\rho(\bm{\mathcal{J}},t) = {\mathbb P}[\bm{\mathcal X}(t) = \bm{\mathcal J}|\mathcal{E}(t)].
\end{equation}
The belief is recursively computed by exploiting the conditional independence of the hidden Markov chain, observing that $\mathcal{E}(t) = \{\bm{\beta}(t), \mathcal{E}(t-1)\}$ and setting the initial system conditions as
\begin{equation}
    \rho(\bm{\mathcal{I}}, 0)=
    \begin{cases}
        1 \quad \bm{\mathcal{I}} = \bm{0},\\
        0 \quad \text{otherwise}.  
    \end{cases}    
\end{equation}
The belief at the end of the \ac{pia} sub-frame of frame $t$ is computed as 
\begin{equation}\label{posteriorprob}
\rho(\bm{\mathcal{J}}, t) = e_{\bm{\beta}}(\bm{\mathcal{J}}, \bm{a})\sum_{\bm{\mathcal{I}}: \bm{\beta}(t) = \bm{\beta}} p_{\bm{\mathcal{I}}\bm{\mathcal{J}}}(\bm{q}(t-1))\rho(\bm{\mathcal{I}}, t-1).
\end{equation} 
Let also $\bm{\rho}(t)$ be the vector collecting  $\rho(\bm{\mathcal{J}},t)$ for all $\bm{\mathcal{J}}$.

\vspace{5pt}\noindent\emph{Penalty.} When moving from state $\bm{\mathcal I}$ to state $\bm{\mathcal J}$ with action $\bm{a}$, the penalty is $P(\bm{\mathcal{I}}, \bm{\mathcal{J}}, \bm{a})$, where the penalty function depends only on the states and action.

\subsection{Latency As Penalty Function}

As a performance metric, we consider the packet latency, which measures the time a packet spends in the buffer before successful transmission. Such time includes the delay introduced by collisions and retransmissions. 

In formulas, the \textit{latency} of packet $h$ in user $n$ buffer is defined as
\begin{equation}
    D_{n, h} = \tau_{n,h} - x_{n, h},
\end{equation}
where $\tau_{n,h}$ is the instant of the successful delivery of the packet to the \ac{bs}.
Note that all packets in user $n$ buffer at time $\tau_0(t)$ increase their latency by $L(t) = L_1 + L_2(t)$ if they are not successfully transmitted at frame $t$. Instead, a successful transmission provides an increment of $L_1 +q_n(t)$. 

We consider the latency increment penalty associated with action $\bm{a}$ when in state $\bm{\mathcal{I}}$ at frame $t$, defined as 
\begin{equation}
    P(\bm{\mathcal I}, \bm{\mathcal J}, \bm{q}(t)) = \sum_{n = 1}^{N}\sum_{h = 1}^{K_n(\tau_0(t))}d_{n, h}(t),
\end{equation}
where the incremental latency of packet $h$ in user $n$ buffer is 
\begin{equation}
    d_{n, h}(t) =
    \begin{cases}
        L_1 + q_n(t-1), &  h=1,\;z_n(t-1) = 1,\\
        L(t-1), &  z_n(t-1) = 0,\\
        \tau_0(t) - x_{n,h}& \text{otherwise},
    \end{cases}
\end{equation}
and we note that from the couple of subsequent states $\bm{\mathcal I}, \bm{\mathcal J}$ at frames $t$ and $t+1$ we can compute $\bm{z}(t)$. Since we consider \ac{fifo} buffers, $h = 1$ refers to the oldest packet in the buffer, which is also the first scheduled for transmission in the buffer.

\subsection{\ac{pomdp} Solution}
As for conventional \acp{mdp}, the solution of the \ac{pomdp} minimizes the expected future penalty $\mathbb{E}\left[\sum _{t=0}^{\infty }P(t)\right]$. Note that, since the agent does not directly observe the environment's state, it must make decisions based on its belief of the current state.
For this reason, the \ac{pomdp} is typically formulated as a conventional infinite-states \ac{mdp} by moving from space $\mathcal{Y}$ to the continuous space of all distributions of the beliefs, thus all possible values of $\rho(\bm{\mathcal{J}})$, for all $\bm{\mathcal{J}} \in \mathcal{Y}$.
Both the observations and actions are the same of the original \ac{pomdp}, and the \textit{policy} function maps each belief state to an action,
\begin{equation}
\pi: \bm{\rho}(t) \rightarrow \bm{q}(t).
\end{equation}
The optimal policy, i.e, the one minimizing the long-term penalty, is the solution of the Bellman equation applied to the belief \ac{mdp}, as detailed in \cite{Kaelbling98}.

\section{Frame-Efficiency Optimization}\label{sec:frameeff}
The direct \ac{pomdp} solution is, in general, very challenging, due to the infinite number of hidden states of the \ac{pomdp} model. Therefore, we consider a sub-optimal solution that determines the frame length $L_2(t)$ iteratively, while performing the user assignments. 
To this end, we first resort to the frame efficiency metric \cite{rech2023partial}, which can be extended to the correlated activation case exploiting the \ac{pomdp} formulation.

Let $l \in \{1,\ldots, L_2(t)\}$ be the slot index within frame $t$ (in the \ac{dt} sub-frame), and let $c_l = 1$ if a successful transmission occurs at slot $l$ and $c_l = 0$ otherwise.
The \textit{conditional frame efficiency} is defined as 
\begin{equation}\label{frameeff}
    \eta(t) = \frac{1}{L_2(t)}\sum_{l=1}^{L_2(t)}\mathbb{E}[c_l|\mathcal{E}(t)], 
\end{equation}
where, with respect to the definition in \cite{rech2023partial}, the condition on $\mathcal{E}(t)$ accounts for the transmission outcomes history.

At frame $t$, immediately after the end of the \ac{pia} sub-frame, the \ac{bs} solves the following optimization problem:
\begin{subequations}\label{feopt}
	\label{maxprob}
	\begin{equation}
\max_{\substack{\bm{q}(t)}}\eta(t),
	\end{equation}
	\begin{equation}
{\rm     s.t.\;}		\;\; q_n(t) \in \{1,\ldots, L_2(t)\}.
	\end{equation}
\end{subequations}
Still, problem \eqref{feopt} is one of \ac{minlp}, and its solution quickly becomes for large numbers of users. Therefore, to simplify the analysis, we resort to a greedy scheduling solution.

\subsection{Greedy Frame Efficiency Optimizer Algorithm}\label{sec:GFEO}
 
The proposed algorithm is denoted as \ac{gfeo} algorithm, and it iteratively adjusts the frame length $L_2(t)$, while performing the user assignments $\bm{q}(t)$. 
First, note that, due to the assumption of buffering packets generated during frame $t$ and single slot assignment to each user, each hidden state/action couple leads to a single possible observation $\bm{\beta}^*$. Thus, the emission probability is $e_{\bm{\beta}}(\bm{\mathcal{J}},\bm{a}) = 1$, for $\bm{\beta} = \bm{\beta}^*$, and  $e_{\bm{\beta}}(\bm{\mathcal{J}},\bm{a}) = 0$ otherwise.
Hence, \eqref{posteriorprob} can be rewritten and recursively computed as
\begin{equation}\label{posteriorprob_simplified}
\rho(\bm{\mathcal{J}}, t) = \sum_{\bm{\mathcal{I}}: \bm{\beta}(t) = \bm{\beta}^*} p_{\bm{\mathcal{I}}\bm{\mathcal{J}}}(\bm{q}(t-1))\rho(\bm{\mathcal{I}}, t-1).
\end{equation}
Now, let $\eta(L_2(t), \bm{q}(t)) = \eta(t)$ be the expected conditional frame efficiency, where we now highlight its dependence on the \ac{dt} sub-frame length and the user assignment. Let us also introduce the activation probability of user $n$ conditioned to the current belief states, i.e., from \eqref{posteriorprob_simplified}, 
\begin{equation}\label{actprob}
    \phi_n(t) = \sum_{\bm{\mathcal{I}} : K_n(\tau_0(t))>0}\rho(\bm{\mathcal{I}}, t).
\end{equation}
Finally suppose, without loss of generality, that users' indices are ordered with decreasing activation probability, i.e.,
\begin{equation}
\phi_1(t)\geq \phi_2(t)\geq \ldots \geq \phi_N(t). 
\end{equation}

The \ac{gfeo} algorithm includes $N$ iterations (one per user), assigning user $n$ to a specific time slot at iteration $n$. At the first iteration ($n=1$), user 1 is assigned to the first \ac{dt} slot of the frame, i.e., $q_1(t) = 1$. 

The algorithm compares the frame efficiency obtained by assigning the current user to each of the already allocated slots, or to a new slot, thus increasing $L_2(t)$. Then, the best solution among those explored is chosen, and the algorithm moves to the scheduling of the next user.
Specifically, at iteration $n\geq2$, the algorithm computes the conditional frame efficiency obtained by assigning user $n$ to each slot $l \in \{1,\ldots, L_2(t)\}$.
Letting $q_m(t)$, $m=1, \ldots, n-1$, be the indices of the slots assigned to the  previous $n-1$ users, the \ac{gfeo} assigns user $n$ to the slot with index
\begin{equation}
\begin{split}
    q_n(t) = \argmax_{\ell}\left\{\max\left[\eta(L_2(t), \tilde{\bm{q}}_{n}(l,t)), \right.\right.\\
    \left.\left.  \eta(L_2(t)+1, \tilde{\bm{q}}_{n}(l,t))\right]\right\},
\end{split}
\end{equation}
where $\tilde{\bm{q}}_{n}(l,t) = [q_1(t), \ldots, q_{n-1}(t), l, 0, \ldots, 0]$.
The frame length $L_2(t)$ is updated at each iteration, depending on the frame efficiency provided by each assignment.
Note that the \ac{bs} exploits only the partial information acquired through the observation history, while the instantaneous state of the hidden Markov process is never exploited, as the \ac{bs} has no knowledge of the buffers conditions. Thus, in general, this algorithm provides a sub-optimal policy.

\vspace{5pt}\noindent\emph{Practical \ac{mdp} Considerations.}
Due to the potentially infinite state space provided by the arrival instants of the packets in the users' buffers, it is impossible to compute \eqref{actprob} in practice. As a first approximation, instead of considering the state of the arrival instants, we consider the state of the buffers load, i.e., the state is described by the number of packets in each user's buffer.
Still, under the assumption of infinite buffer capacities, the state space remains infinite. Nevertheless, we can assume finite buffers of length $C$ and analyze the system in stability conditions, wherein the number of packets in each user is always less than $C$ in practice.
Under this further assumption, the total number of \ac{pomdp} states is $|\mathcal{Y}| = (C+1)^{N}$.

\subsection{Simplified \ac{gfeo}}
Although being extremely accurate in the estimation of the users' buffer states and their correlation, due to the memory of the previous observations, the \ac{pomdp} model quickly becomes unfeasible with the increasing number of users in the system. In particular, as the number of states increases exponentially with $N$, the complexity of the computation of the posterior probabilities \eqref{posteriorprob_simplified} drastically increases.

To overcome this limitation, we further simplify the state space and consider a low-complexity variant of the proposed greedy scheduler, named \ac{sgfeo}.
With \ac{sgfeo}, at frame $t$ the recursion in \eqref{posteriorprob_simplified} is removed by considering $\mathcal{E}(t) = \{\bm{\beta}(t), \nu(t-1)\}$, i.e., only the observation of frame $t-1$ and not the entire history of actions and observations.
Under this assumption, the memory of the state transitions is not kept, and the computation of $\rho(\bm{\mathcal{I}}, t-1)$ is avoided by setting a uniform distribution of the states compatible with the current observation $\mathcal{E}(t)$.
Let $\mathcal{Z}(t) = \{\bm{\mathcal{I}}: \mathcal{E}(t) = \bm{\beta}^*\}$ be the set of state compatible with observation $\bm{\beta}^*$, the posterior probabilities at frame $t$ are computed from \eqref{posteriorprob} and \eqref{posteriorprob_simplified} as 
\begin{equation}\label{posteriorprob_sgfeo}
\rho(\bm{\mathcal{J}}, t) = \sum_{\bm{\mathcal{I}} \in \mathcal{Z}} p_{\bm{\mathcal{I}}\bm{\mathcal{J}}}(\bm{q}(t-1))\frac{1}{|\mathcal{Z}(t)|}.
\end{equation}
Then, \ac{sgfeo} proceeds with the users' scheduling as \ac{gfeo}.

\section{Numerical Results}\label{sec:numericalresults}
In this section, we present the numerical results, comparing the \ac{gfeo} and \ac{sgfeo} schedulers with a) the \ac{tdma}, providing fixed-duration frames of $N$ slots with one user assigned per slot deterministically, b) the Rivest's stabilized \ac{saloha} \cite{Rivest87Network}, wherein users generating packets are backlogged with the same probability, and c) the \ac{pima} protocol of \cite{rech2023partial}.
For all schemes, we adopt the third numerology of the \ac{nr} specification, which provides \ac{dt} time slots of $T_{\rm s} = 0.125$~ms \cite{3GPP38211}. Moreover, for \ac{pima}, \ac{gfeo}, and \ac{sgfeo}, we assume an error-free user enumeration with a fixed $L_1$. 

The performance is assessed in terms of \textit{average frame efficiency} $\bar{\eta}$ (conditioned on $\nu(t)>0$), and \textit{average latency} $\bar{D} = \mathbb{E}[D_{n, h}]$, computed from all successfully delivered packets.

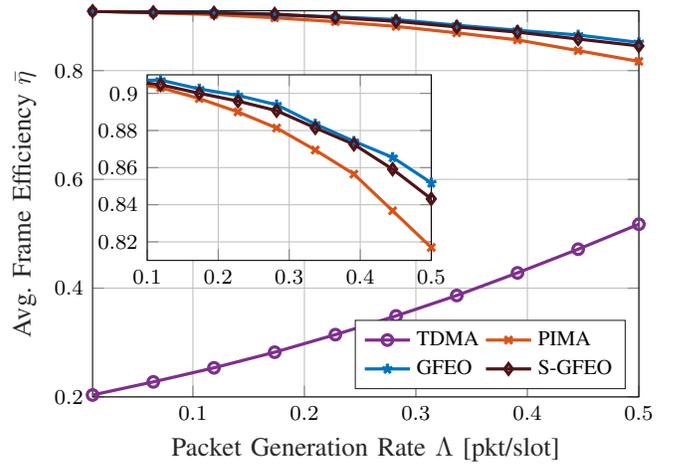
\begin{figure}
    \centering
    \setlength\fwidth{0.82\columnwidth}
    \setlength\fheight{0.58\columnwidth}
    \definecolor{TDMAcolor}{rgb}{0.49400,0.18400,0.55600}%
\definecolor{PIMAcolor}{rgb}{0.85000,0.32500,0.09800}%
\definecolor{GFEOcolor}{rgb}{0.00000,0.44700,0.74100}%
\definecolor{sGFEOcolor}{rgb}{0.3,0.08,0.09}%
\pgfplotsset{every tick label/.append style={font=\footnotesize}}

\begin{tikzpicture}

\begin{axis}[%
    width=\fwidth,
    height=\fheight,
    at={(0\fwidth,0\fheight)},
    scale only axis,
    ylabel style={font=\normalsize},
    xlabel style={font=\normalsize},
    xmin=0.01,
    xmax=0.5,
    xlabel style={font=\color{white!15!black}},
    xlabel={Packet Generation Rate  $\Lambda$ [pkt/slot]},
    ymin=0.2,
    ymax=0.91,
    yminorticks=true,
    ylabel style={font=\color{white!15!black}},
    ylabel={Avg. Frame Efficiency $\bar{\eta}$},
    axis background/.style={fill=white},
    xmajorgrids,
    ymajorgrids,
    legend style={at={(0.48, 0.02)}, legend columns = 2, anchor=south west, legend cell align=left, align=left, font=\footnotesize, draw=white!15!black}
]

\addplot [color=TDMAcolor, very thick, mark size=2.0pt, mark=o, mark options={solid, TDMAcolor}]
  table[row sep=crcr]{%
  0.01	0.203503649635036\\
0.0644444444444444	0.22759226096422\\
0.118888888888889	0.253445420965484\\
0.173333333333333	0.282163424881981\\
0.227777777777778	0.314428929466522\\
0.282222222222222	0.348883889946355\\
0.336666666666667	0.386471235893199\\
0.391111111111111	0.428058756074917\\
0.445555555555556	0.471581036430237\\
0.5	0.517524688109464\\
};
\addlegendentry{TDMA}

\addplot [color=PIMAcolor, very thick, mark size=2.0pt, mark=x, mark options={solid, PIMAcolor}]
  table[row sep=crcr]{%
  0.01	0.909099783646055\\
0.0644444444444444	0.90614299502672\\
0.118888888888889	0.903063592214821\\
0.173333333333333	0.897202299109906\\
0.227777777777778	0.890025162014186\\
0.282222222222222	0.881348520366154\\
0.336666666666667	0.869484919598562\\
0.391111111111111	0.856509641621457\\
0.445555555555556	0.836743244470565\\
0.5	0.816969632170478\\
};
\addlegendentry{PIMA}

\addplot [color=GFEOcolor, very thick, mark size=2.0pt, mark=star, mark options={solid, GFEOcolor}]
  table[row sep=crcr]{%
0.01	0.909090909090989\\
0.0644444444444444	0.906966854351758\\
0.118888888888889	0.90714772870157\\
0.173333333333333	0.902394594332535\\
0.227777777777778	0.898903753953413\\
0.282222222222222	0.893858388894202\\
0.336666666666667	0.883506214602496\\
0.391111111111111	0.874187401764232\\
0.445555555555556	0.865528455855568\\
0.5	0.851797885216885\\
};
\addlegendentry{GFEO}

\addplot [color=sGFEOcolor, very thick, mark size=2.0pt, mark=diamond, mark options={solid, sGFEOcolor}]
  table[row sep=crcr]{%
0.01	0.909090909090988\\
0.0644444444444444	0.906966854351758\\
0.118888888888889	0.905998633544456\\
0.173333333333333	0.903974052045644\\
0.227777777777778	0.89759865121331\\
0.282222222222222	0.890666444826062\\
0.336666666666667	0.880285324201327\\
0.391111111111111	0.870792452861302\\
0.445555555555556	0.858094216850975\\
0.5	0.84541622542432\\
};
\addlegendentry{S-GFEO}

\end{axis}

\begin{axis}[%
width=0.52\fwidth,
height=0.34\fwidth,
at={(0.1\fwidth,0.25\fwidth)},
scale only axis,
xmin=0.1,
xmax=0.5,
xlabel style={font=\color{white!15!black}},
ymin=0.81,
ymax=0.91,
xtick={0.1, 0.2, 0.3, 0.4, 0.5},
axis background/.style={fill=white},
xmajorgrids,
ymajorgrids
]

\addplot [color=PIMAcolor, very thick, mark size=2.0pt, mark=x, mark options={solid, PIMAcolor}]
  table[row sep=crcr]{%
  0.01	0.909099783646055\\
0.0644444444444444	0.90614299502672\\
0.118888888888889	0.903063592214821\\
0.173333333333333	0.897202299109906\\
0.227777777777778	0.890025162014186\\
0.282222222222222	0.881348520366154\\
0.336666666666667	0.869484919598562\\
0.391111111111111	0.856509641621457\\
0.445555555555556	0.836743244470565\\
0.5	0.816969632170478\\
};
\addlegendentry{PIMA}

\addplot [color=GFEOcolor, very thick, mark size=2.0pt, mark=star, mark options={solid, GFEOcolor}]
  table[row sep=crcr]{%
0.01	0.909090909090989\\
0.0644444444444444	0.906966854351758\\
0.118888888888889	0.90714772870157\\
0.173333333333333	0.902394594332535\\
0.227777777777778	0.898903753953413\\
0.282222222222222	0.893858388894202\\
0.336666666666667	0.883506214602496\\
0.391111111111111	0.874187401764232\\
0.445555555555556	0.865528455855568\\
0.5	0.851797885216885\\
};
\addlegendentry{GFEO}

\addplot [color=sGFEOcolor, very thick, mark size=2.0pt, mark=diamond, mark options={solid, sGFEOcolor}]
  table[row sep=crcr]{%
0.01	0.909090909089883\\
0.0644444444444444	0.907172182216927\\
0.118888888888889	0.904629141119359\\
0.173333333333333	0.900047841880943\\
0.227777777777778	0.895836429542401\\
0.282222222222222	0.890676461074109\\
0.336666666666667	0.881419767033108\\
0.391111111111111	0.87253346961074\\
0.445555555555556	0.859155354265282\\
0.5	0.843160957413455\\
};
\addlegendentry{S-GFEO} 

\legend{}
\end{axis}

\end{tikzpicture}
    \vspace{-10pt}
     \caption{Average frame efficiency versus the total arrival rate for $N = 5$. The zoom plot inside the figure highlights the performance gap between the \ac{pima}-based schedulers.}
     \label{fig:frameeffN5}
\end{figure}

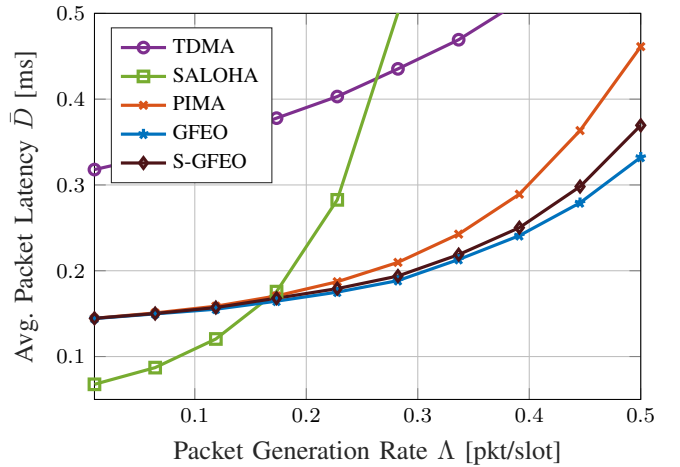
\begin{figure}
    \centering
    \setlength\fwidth{0.82\columnwidth}
    \setlength\fheight{0.58\columnwidth}
    \definecolor{SAL0HAcolor}{rgb}{0.46600,0.67400,0.18800}%
\definecolor{TDMAcolor}{rgb}{0.49400,0.18400,0.55600}%
\definecolor{PIMAcolor}{rgb}{0.85000,0.32500,0.09800}%
\definecolor{GFEOcolor}{rgb}{0.00000,0.44700,0.74100}%
\definecolor{sGFEOcolor}{rgb}{0.3,0.08,0.09}%
\pgfplotsset{every tick label/.append style={font=\footnotesize}}

\begin{tikzpicture}

\begin{axis}[%
    width=\fwidth,
    height=\fheight,
    at={(0\fwidth,0\fheight)},
    scale only axis,
    ylabel style={font=\normalsize},
    xlabel style={font=\normalsize},
    xmin=0.01,
    xmax=0.5,
    xlabel style={font=\color{white!15!black}},
    xlabel={Packet Generation Rate $\Lambda$ [pkt/slot]},
    ymin=0.05,
    ymax=0.5,
    yminorticks=true,
    ylabel style={font=\color{white!15!black}},
    ylabel={Avg. Packet Latency $\bar{D}$ [ms]},
    axis background/.style={fill=white},
    xmajorgrids,
    ymajorgrids,
    legend style={at={(0.03,0.56)}, anchor=south west, legend cell align=left, align=left, font=\footnotesize, draw=white!15!black}
]

\addplot [color=TDMAcolor, very thick, mark size=2.0pt, mark=o, mark options={solid, TDMAcolor}]
  table[row sep=crcr]{%
  0.01	0.31792488939971\\
0.0644444444444444	0.333774211164372\\
0.118888888888889	0.35313673313179\\
0.173333333333333	0.377851934369687\\
0.227777777777778	0.403029520094215\\
0.282222222222222	0.435289584360361\\
0.336666666666667	0.469249616601617\\
0.391111111111111	0.513541140861119\\
0.445555555555556	0.564845166256827\\
0.5	0.627217261959476\\
};
\addlegendentry{TDMA}

\addplot [color=SAL0HAcolor, very thick, mark size=2.0pt, mark=square, mark options={solid, SAL0HAcolor}]
  table[row sep=crcr]{%
  0.01	6.76430693546184e-02\\
0.0644444444444444	8.71067216556092e-02\\
0.118888888888889	0.120483535155444\\
0.173333333333333	0.175671665952376\\
0.227777777777778	0.282629368117388\\
0.282222222222222	0.500671121862125\\
0.336666666666667	1.17459015012754\\
0.391111111111111	3.36181838377575\\
0.445555555555556	5.78166445965338\\
0.5	6.51927741873411\\
};
\addlegendentry{SALOHA}

\addplot [color=PIMAcolor, very thick, mark size=2.0pt, mark=x, mark options={solid, PIMAcolor}]
  table[row sep=crcr]{%
0.01	0.144535635802063\\
0.0644444444444444	0.150824840109334\\
0.118888888888889	0.158682487043781\\
0.173333333333333	0.170674594084557\\
0.227777777777778	0.187176143869526\\
0.282222222222222	0.209753397835066\\
0.336666666666667	0.242647888321656\\
0.391111111111111	0.289210501250792\\
0.445555555555556	0.363385967142099\\
0.5	0.461221879823604\\
};
\addlegendentry{PIMA}

\addplot [color=GFEOcolor, very thick, mark size=2.0pt, mark=star, mark options={solid, GFEOcolor}]
  table[row sep=crcr]{%
0.01	0.14428150060133\\
0.0644444444444444	0.149737068738054\\
0.118888888888889	0.15517966089862\\
0.173333333333333	0.164542520685278\\
0.227777777777778	0.175082894639887\\
0.282222222222222	0.188517272963222\\
0.336666666666667	0.213093623489578\\
0.391111111111111	0.240916916915743\\
0.445555555555556	0.279197982250964\\
0.5	0.332225348573691\\
};
\addlegendentry{GFEO}

\addplot [color=sGFEOcolor, very thick, mark size=2.0pt, mark=diamond, mark options={solid, sGFEOcolor}]
  table[row sep=crcr]{%
0.01	0.144552170840922\\
0.0644444444444444	0.150139460219197\\
0.118888888888889	0.157123917210172\\
0.173333333333333	0.168019647088434\\
0.227777777777778	0.179158350855206\\
0.282222222222222	0.193843686287554\\
0.336666666666667	0.218795216925745\\
0.391111111111111	0.250194536690086\\
0.445555555555556	0.298114157499382\\
0.5	0.369490269017218\\
};
\addlegendentry{S-GFEO}

\end{axis}
\end{tikzpicture}
    \vspace{-10pt}
     \caption{Average packet latency versus the packet arrival rate for $N = 5$.}
     \label{fig:latencyN5}
     \vspace{-10pt}
\end{figure}

Firstly, we consider a scenario with $N = 5$ users with traffic intensity $0.01\leq\Lambda\leq0.5$, which corresponds to the queues' stability regime of the \ac{gfeo} algorithm, which has been derived empirically from the simulations. The length of the \ac{pia} sub-frame is set to $L_1 = T_{\rm s}/10$, which is sufficient to guarantee negligible enumeration error probability \cite{rech2023partial}.
Fig.~\ref{fig:frameeffN5} shows the average frame efficiency $\bar{\eta}$ as a function of the total packet generation rate. 
Note that the performance of \ac{saloha} is not reported, as the frame efficiency cannot be defined for frame-less protocols \ac{saloha}.
\Ac{tdma}, adopting the constant maximum frame length, provides a very low frame efficiency, while the \ac{pima}-based schedulers attain close-to-optimal performance in the whole considered range of traffic intensity. Among these schedulers, \ac{gfeo}, which exploit the correlation knowledge at best, achieve the highest frame efficiency providing approximately up to 5\% gain with respect \ac{pima}. However, the results also show that the gap between \ac{gfeo} and \ac{sgfeo} is almost negligible, therefore the approximations introduced with \ac{sgfeo} have a very limited negative impact on the system performance.  
Finally, while the frame efficiency is increasing with the traffic intensity for \ac{tdma} due to the reduced unused time slots, it is slightly decreasing for the \ac{pima}-based schemes due to the increasing chances of collisions.

The comparison of the average packet latency is shown in Fig.~\ref{fig:latencyN5}.
At low traffic, \ac{tdma} provides the larger latency due to the frame length fixed to $N$, while \ac{saloha} achieves extremely low latency due to the absence of collisions. Instead, at high traffic intensity, the \ac{saloha} backlogging mechanism prevents the users to transmit their buffered packets immediately, therefore increasing the average latency. The \ac{pima}-based schemes, due to their better capability of adapting to traffic conditions, outperform both schemes when the traffic intensity is moderately large. In particular, both \ac{gfeo} and \ac{sgfeo} achieve the lowest latency, still with an almost negligible gap between these greedy solutions.
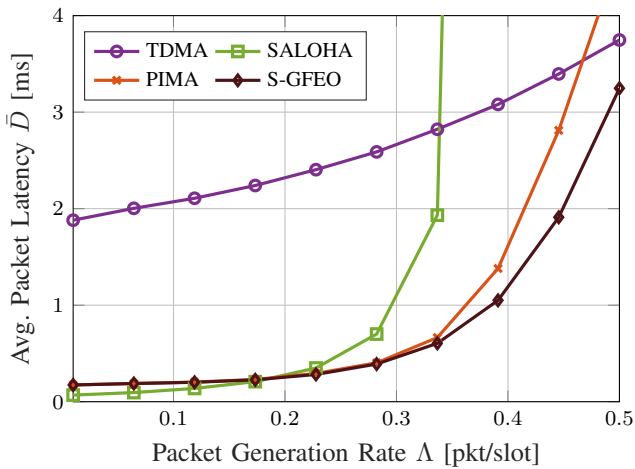
\begin{figure}
    \centering
    \setlength\fwidth{0.82\columnwidth}
    \setlength\fheight{0.58\columnwidth}
    \definecolor{SAL0HAcolor}{rgb}{0.46600,0.67400,0.18800}%
\definecolor{TDMAcolor}{rgb}{0.49400,0.18400,0.55600}%
\definecolor{PIMAcolor}{rgb}{0.85000,0.32500,0.09800}%
\definecolor{GFEOcolor}{rgb}{0.00000,0.44700,0.74100}%
\definecolor{sGFEOcolor}{rgb}{0.3,0.08,0.09}%
\pgfplotsset{every tick label/.append style={font=\footnotesize}}

\begin{tikzpicture}

\begin{axis}[%
    width=\fwidth,
    height=\fheight,
    at={(0\fwidth,0\fheight)},
    scale only axis,
    ylabel style={font=\normalsize},
    xlabel style={font=\normalsize},
    xmin=0.01,
    xmax=0.5,
    xlabel style={font=\color{white!15!black}},
    xlabel={Packet Generation Rate $\Lambda$ [pkt/slot]},
    ymin=0,
    ymax=4,
    yminorticks=true,
    ylabel style={font=\color{white!15!black}},
    ylabel={Avg. Packet Latency $\bar{D}$ [ms]},
    axis background/.style={fill=white},
    xmajorgrids,
    ymajorgrids,
    legend style={at={(0.02,0.78)}, legend columns = 2, anchor=south west, legend cell align=left, align=left, font=\footnotesize, draw=white!15!black}
]

\addplot [color=TDMAcolor, very thick, mark size=2.0pt, mark=o, mark options={solid, TDMAcolor}]
  table[row sep=crcr]{%
0.01	1.88059071294002\\
0.0644444444444444	2.00341679985934\\
0.118888888888889	2.10764506521103\\
0.173333333333333	2.23984321854857\\
0.227777777777778	2.40390866453411\\
0.282222222222222	2.58804043192765\\
0.336666666666667	2.82270740113768\\
0.391111111111111	3.07996782246917\\
0.445555555555556	3.39552926502713\\
0.5	3.74754991265324\\
};
\addlegendentry{TDMA}

\addplot [color=SAL0HAcolor, very thick, mark size=2.0pt, mark=square, mark options={solid, SAL0HAcolor}]
  table[row sep=crcr]{%
0.01	6.85312970286363e-02\\
0.0644444444444444	9.48270923383971e-02\\
0.118888888888889	0.137607179204543\\
0.173333333333333	0.207734419452893\\
0.227777777777778	0.349116682465531\\
0.282222222222222	0.701011377467967\\
0.336666666666667	1.93247271620439\\
0.391111111111111	27.1129377251512\\
0.445555555555556	38.0875073919692\\
0.5	41.3398422759061\\
};
\addlegendentry{SALOHA}

\addplot [color=PIMAcolor, very thick, mark size=2.0pt, mark=x, mark options={solid, PIMAcolor}]
  table[row sep=crcr]{%
0.01	0.173390310120278\\
0.0644444444444444	0.188406113754686\\
0.118888888888889	0.203258417437093\\
0.173333333333333	0.230531417782098\\
0.227777777777778	0.292969253581505\\
0.282222222222222	0.402974159130462\\
0.336666666666667	0.664677989729387\\
0.391111111111111	1.38027228894486\\
0.445555555555556	2.81097287507392\\
0.5	4.64028336070853\\
};
\addlegendentry{PIMA}

\addplot [color=sGFEOcolor, very thick, mark size=2.0pt, mark=diamond, mark options={solid, sGFEOcolor}]
  table[row sep=crcr]{%
0.01	0.173627218312221\\
0.0644444444444444	0.188506468611375\\
0.118888888888889	0.201185417437039\\
0.173333333333333	0.226537870921418\\
0.227777777777778   0.282356925888150\\
0.282222222222222	0.389973046241591\\
0.336666666666667	0.604678938972914\\
0.391111111111111	1.050289724828464\\
0.445555555555556	1.910903972787521\\
0.5	3.24708530283360\\
};
\addlegendentry{S-GFEO}

\end{axis}
\end{tikzpicture}   
    \caption{Average packet latency versus the packet arrival rate for $N = 30$.}
    \label{fig:latencyN30}
    \vspace{-5pt}
\end{figure}

Finally, we consider a scenario with $N = 30$ users and a fixed \ac{pia} sub-frame length $L_1 = T_{\rm s}/4$. Here, the performance of \ac{gfeo} is not reported due to its extremely high complexity for a large number of users. However, we have observed that its low-complexity variant \ac{sgfeo} scales better at the cost of a negligible loss in performance.
Fig.~\ref{fig:latencyN30} depicts the average packet latency achieved by the compared schemes. First, note that similarly to Fig.~\ref{fig:latencyN5}, the results show a huge latency reduction of the \ac{pima}-based solutions with respect to the other considered schenes. 
We stress that the average latency counts only for the successfully delivered packets and that we are considering values of $\Lambda$ such that the queues' stability is verified for \ac{pima} and \ac{gfeo}. 
Therefore, while \ac{tdma} seems to outperform \ac{pima} at high traffic, this is only due to the instability conditions of \ac{tdma}, which drops the oldest packets in the users' queues when the maximum capacity is reached.
Still, we observe that \ac{sgfeo} achieves the lowest latency among all the compared schemes.
\balance 

\section{Conclusions}\label{sec:conclusions}
To minimize the latency of our recently proposed \ac{pima} scheme, we proposed to model the protocol with a \ac{pomdp}.
The scheduling is optimized based on the knowledge of both the number of active users and past transmission outcomes, which are mapped to the observation of a hidden Markov process. To overcome the huge complexity of the scheduling problem, we proposed two sub-optimal greedy schedulers namely \ac{gfeo} and \ac{sgfeo}, both aiming at dynamically maximizing the frame efficiency. Numerical results show the effectiveness of both \ac{gfeo} and \ac{sgfeo} in attaining extremely low latency over a wide range of traffic intensity.

\bibliographystyle{IEEEtran}
\bibliography{IEEEabrv, Bibliography.bib}

\end{document}